\documentstyle[aps,pra,prabib]{revtex}
\newcommand{\eeqar}[1]{\label{#1} \end{eqnarray}}
\begin{document}
\tightenlines
\title{Quantum Chaos, Degeneracies and Exceptional Points}
\author{W.D.$\,$ Heiss  and S.$\,$ Radu }
\address{
Centre for Nonlinear Studies and Department of Physics\\
University of the Witwatersrand, PO Wits 2050, Johannesburg, South Africa }
\maketitle
\begin{abstract}
It is argued that, if a regular Hamiltonian is perturbed by a term
that produces chaos, the onset of chaos is shifted towards larger values of
the perturbation parameter if the unperturbed spectrum
is degenerate and the lifting of the degeneracy is of second order in this
parameter. The argument is based on the behaviour of the
exceptional points of the full problem.
\end{abstract}
\vskip 1cm
PACS Nos.:  36.40.+d, 05.45.+b
\section{Introduction}
The study of single particle motion in deformed mean fields has attracted
much attention recently because of its relevance to nuclear physics and for
the description of metallic clusters \cite{ba,henara}.
For the case of harmonic oscillator
potentials, deformations which go beyond a quadrupole deformation are
nonintegrable and show chaotic behaviour in the classical and
quantum mechanical treatment. However, closer scrutiny has revealed that
the addition of an octupole term to a prolate quadrupole potential produces
the typical signatures of chaos only for fairly large octupole strength,
in fact, the problem appears to be close to
integrability. Moreover, the quantum mechanical treatment produces new shell
structures for finite octupole strength even though the classical problem
is nonlinear \cite{henara,arit}.
In contrast, addition of an octupole term to an oblate
quadrupole deformed potential yields chaos with a positive Lyapunov exponent,
and the corresponding quantum spectrum has the typical statistical properties
ascribed to quantum chaos.

In this paper we address the question: what are the intrinsic properties
of the quantum mechanical operators which give rise to the different
behaviour described above. This question is of interest for a possible
characterisation of what is called quantum chaos. The example mentioned
in the previous
paragraph renders a good case for such studies as it refers to the
orthodox situation where a classically chaotic system is treated quantum
mechanically. Our aim is to unravel the universal operator properties that
produce the typical patterns ascribed to quantum chaos without reference
to an underlying classical system. There is progress in previous work
towards this aim as it has been recognised that, for a problem of the
form $H^0+\lambda H^1$, it is precisely the high density of the exceptional
points \cite{ben,heste}, i.e.$\,$ the singularities of the spectrum
$E_n(\lambda )$,
which bring about the statistical properties of the spectrum associated with
quantum chaos \cite{hesan}.
The new aspect in the present paper is the effect of
degeneracies at $\lambda =0$ upon the behaviour for $\lambda >0$.
We find that if the
lifting of degeneracies is a second order effect in $\lambda$,
then the onset
of chaos is suppressed and will occur only for larger values of $\lambda$.
It is
this pattern which explains the difference between the prolate and oblate
case as described above. The major point of the present paper is the
universal validity of this finding, and the behaviour of the exceptional
points around $\lambda =0$ provides for the explanation.

In the following section we introduce briefly the concept of exceptional
points and present the argument for the statement made above. Section three
contains examples for illustration and section four refers to the particular
physical example introduced above. A summary and discussion is given in the
last section.

\section{Effect of Degeneracies on Exceptional Points}
To make the paper selfcontained
we briefly recapitulate the significance of exceptional points and their
connection to avoided level crossings. This will facilitate the discussion
about the effect of degeneracies on exceptional points and hence the global
structure of the spectrum.

There is essentially a one to one relationship between avoided level
crossings and exceptional points \cite{hesan}. If we have a quantum
mechanical
problem of the form $H^0+\lambda H^1$
with $H^0$ and $H^1$ given as $N \times N$
matrices, then the spctrum $E_n(\lambda )\, ,n=1,\ldots ,N$ is
determined by one
analytic function evaluated on $N$ Riemann sheets \cite{heste}. If there are
no degeneracies, the $N$ sheets are connected by $N(N-1)$ branch points, the
exceptional points. They occur in complex conjugate pairs and are the points
in the complex $\lambda $-plane where
any two pairs of energies coalesce.  If this
happens sufficiently close to the real $\lambda $-axis, an avoided level
crossing
occurs since the pair of energies coalescing in the complex plane still
assumes values near to each other on the neighbouring real $\lambda $-axis.
In
principle, the positions of the exceptional points are determined by the
resultant of the secular equation for the spectrum, i.e.$\,$ by the
simultaneous solution of the two ploynomial equations
\begin{eqnarray} \det (E-H^0-\lambda H^1)&=&0  \cr
{{\rm d} \over {\rm d} E} \det (E-H^0-\lambda H^1)&=&0
\label{o}
\end{eqnarray}
which leads by elimination of $E$ to the resultant which is
a polynomial of order $N(N-1)$ in $\lambda $.
For large values of $N$, an explicit calculation of the exceptional points
is prohibitive. However, it is possible to determine a distribution
for the real parts of the exceptional points from the knowledge of $H^0$ and
$H^1$ alone \cite{hesan},
and in this way the regions of real $\lambda $-values with large
scale avoided level crossings can be found. These are the regions where
the spectrum shows the properties ascribed to quantum chaos. This method has
been tested and applied to the particular physical situation of the hydrogen
atom in a strong magnetic field \cite{kohe}.

The effect of removing degeneracies by small random perturbation was
discussed previously \cite{hemuch}. In the present paper we address the
particular problem of lifting degeneracies by the term $\lambda H^1$,
i.e.$\,$ we
assume that $H^0$ has certain systematic degeneracies (like for example the
harmonic oscillator). Guided by the particular physical
problem as discussed in the introduction and resumed in section four,
we consider and compare two situations which turn out to be significantly
different in the context of quantum chaotic behaviour. The two situations
are characterised by the presence or absence of a linear term in the
expansion of $E_n(\lambda )$ in powers of $\lambda $ around $\lambda =0$.

It is obvious that the spectrum has a distinctly different behaviour in the
vicinity of $\lambda =0$ for the two cases considered. A non-vanishing linear
term exhibits the typical behaviour of a lifting of degeneracies, when a
perturbation is switched on, in that the levels fan out of the degenerate
level when $\lambda $ is turned on. With a zero linear term the levels stay
closely
together and separate only when the quadratic term becomes significant. From
this picture one intuitively expects that, if the additional term
$\lambda H^1$
produces chaos at all, the onset of chaos is delayed for increasing
$\lambda $
in the latter case when comparing with the former case. In the following we
confirm this expectation using the exceptional points for the argument, while
the next section presents some illustrative examples.

Each $m$-fold degeneracy reduces the total number of exceptional points by
the number $m(m-1)$ if the linear terms are present in the expansion of the
energy levels. If the leading perturbing terms are quadratic, there is a
further reduction by the same number $m(m-1)$. This follows from the resultant
which starts with $\lambda ^{m(m-1)}$ as the lowest order in the former case
and
with $\lambda ^{2m(m-1)}$ in the latter case. This behaviour is valid for each
$m$-fold degeneracy, in other words, if the levels have at $\lambda =0$ the
degeneracies $m_1,m_2,\ldots $, the lowest order terms of the resultant
start with $\lambda ^{m_1(m_1-1)}\cdot \lambda ^{m_2(m_2-1)}\cdots $ when the
lifting
of the degeneracy is linear and with
$\lambda ^{2m_1(m_1-1)}\cdot \lambda ^{2m_2(m_2-1)}\cdots $ for a quadratic
lifting.
A particular $m$-fold degeneracy can be viewed
as a confluent situation where $m(m-1)$ complex conjugate branch points have
merged into the point $\lambda =0$ thus cancelling the singularities
altogether;
likewise, in the quadratic case, $2m(m-1)$ singularities have merged. The
essential point of the argument lies in the comparison between the two cases:
switching on linear terms leads to the emergence of $\sum m_i(m_i-1)$
additional
exceptional points. The distance from $\lambda =0$ of the additional
exceptional
points depends on the magnitude of the linear terms. For typical values they
are rather closer to than remote from the origin. For large matrices the
additional number of exceptional points can amount to many thousands. This
causes avoided level crossings in the spectrum in a region where without the
linear term the spectrum appears smooth.

\section{Illustration of simple examples}
The realistic example which initiated this work produces more than twenty
nine thousand exceptional points for the matrix size used in our work
\cite{henara}. We return to its treatment in the following section. Here
we illustrate the behaviour of the exceptional points in a low dimensional
example. The effect on the spectrum, in particular the different onset of
quantum chaos for the two different cases, is subsequently demonstrated
in a generic matrix model.

We consider a nine dimensional model where $H^0$ has the diagonal entries
(1,1,1,2,2,2,3,3,3). The three-fold degeneracies are lifted by the term
$\lambda H^1$ with
\begin{equation} H^1(\mu )=\pmatrix{\mu &0&0&0.5&2&1.5&2&1.5&0 \cr
0&0&0&1.5&2&3&0.5&1.5&0.5 \cr
0&0&-\mu &2&3&1&2&1.5&2 \cr
0.5&1.5&2&\mu &0&0&1.5&2.5&0 \cr
2&2&3&0&0&0&1&1.5&1 \cr
1.5&3&1&0&0&-\mu &1&1.5&0.5 \cr
2&0.5&2&1.5&1&1&\mu &0&0 \cr
0.5&1.5&1.5&2.5&1.5&1.5&0&0&0 \cr
0&0.5&2&0&1&0.5&0&0&-\mu \cr }
\label{m}
\end{equation}
For $\mu =0$, $H^1$ has only zero entries in the degenerate subspaces of
$H^0$, therefore the nine levels of the full problem
have a vanishing linear term when expanded at $\lambda =0$. Otherwise
the entries of $H^1$ have no significance, they have been generated randomly,
any other choice (with similar orders of magnitude) yields the same
qualitative result. Also, a diagonal form of $H^1$ in the degenerate
subspaces can always be achieved by orthogonal transformations
within the subspaces without changing the global spectrum. The specific form
chosen is convenient as it leaves out additional irrelevant parameters
and it is in line with the physical situation discussed in the following
section.

A nine dimensional model yields 36 exceptional points in
each half plane, our particular case reduces this number to 18. In Fig.(1a)
we display most of the exceptional points in the upper $\lambda $-plane. When
the
parameter $\mu $ is switched on, nine additional exceptional points emerge
from $\lambda =0$ in each half plane. This is illustrated in Fig.(1b). The
basic
difference between the two figures lies in the additional singularities
scattered around $\lambda =0$. Note that the other exceptional points change
only moderately under variation of $\mu $. The corresponding spectra are
displayed in Figs.(2). The growing distance from the origin
of the avoided crossings with increasing value $\mu $ is clearly discernible
in Figs.(2c) and (2d), where only part of the spectrum is shown for better
illustration. These avoided crossings are due to
the 'new' exceptional points. Note that in particular the levels
which originate from the same degenerate energy are affected.
When this occurs on a large scale, the nearest
neighbour distribution of the spectrum will typically assume the Wigner
surmise for rather small values of $\lambda $ while there is no resemblence
to the Wigner curve when the linear terms of the perturbation vanish.
This is demonstrated in the following study case.

Essentially we repeat the model used above. We consider $N>100$ for the
full dimension and the dimensions of the degenerate subspaces can be two,
three, four and so on. The diagonal matrix $H^0$ contains the entries
(1,1,...,2,2,...) and the entries of $H^1$ are filled randomly with elements
between $-1$ and $+1$. The off-diagonal
elements of the diagonal blocks of the degenerate subspaces are set equal to
zero while the corresponding diagonals are filled with the numbers
$\mu (-n_{{\rm deg}}/2+1/2,-n_{{\rm deg}}/2+1,\ldots,n_{{\rm deg}}/2-1/2)$
with $n_{{\rm deg}}$ being the
dimension of the degenerate subspace. Because of the huge number of
exceptional points, we no longer focus our attention on their positions.
Instead, we compare nearest neighbour distributions for $\mu =0$ and
$\mu \ne 0$ for small values of $\lambda $.

In Figs.(3) typical results for the two different cases are presented. For
the same value of $\lambda $ sample averages of ten samples
of nearest neighbour distributions
are illustrated for $\mu =0$ and $\mu =1$. There is of course a continuous
transition from Fig.(3a) to Fig.(3b), and only if the order of magnitude of
the diagonal elements of $H^1$
has reached that of the other matrix elements, that is for $\mu \simeq 1$,
the Wigner distribution has fully developed. We can understand, in terms of
the exceptional points, why for $0<\mu \ll 1$ the situation closely resembles
that of $\mu =0$ even though a large number of exceptional points has already
emerged from the origin of the $\lambda $-plane. When the exceptional points
are still very close to the origin, that is very close to each other, there
is a cancellation with regard to the effect on the spectrum. In fact,
two square root branch points connecting the same Riemann sheets, are barely
noticeable from a distance; the function $z\sqrt{z^2+\varepsilon ^2}$
looks just like
$z^2$ for $|z|\gg \varepsilon $. Only when the branch points have moved out
sufficiently far does the effect on the spectrum become significant as
clearly seen in Figs.(3). The results presented refer to
$n_{{\rm deg}}=8$ and $N=304$, which yields 2128 exceptional points having
emerged from the origin. When $n_{{\rm deg}}$ is increased, the window of
$\lambda $-values, where this transition is strongly pronounced, becomes wider.
Conversely, for smaller values of $n_{{\rm deg}}$ the effect is clearly
seen only in a smaller range of $\lambda $-values. This follows from the
higher density of exceptional points associated with larger values
of  $n_{{\rm deg}}$. Also, the $\lambda $-values beyond which a plain Wigner
distribution occurs even when $\mu =0$ moves closer towards the origin with
increasing $n_{{\rm deg}}$. These two observations can qualitatively be
understood from perturbative arguments in that we expect the onset of a
Wigner distribution beyond the intersection point of two straight lines or
parabolae from neighbouring levels for $\mu \ne 0$ or $\mu =0$, respectively;
the curves intersect at $\lambda _{{\rm intsct}}$ which is proportional
to $1/n_{{\rm deg}}$ in the
former and $1/\sqrt{n_{{\rm deg}}}$ in the latter case. This
argument indicates where perturbation breaks down; we recall that the
present paper deals with a nonperturbative situation as it is just
the exceptional points which break down perturbation. We note that in
a typical physical situation, like the one discussed in
the next section, one is usually faced with steadily increasing values of
$n_{{\rm deg}}$ with increasing energy.

\section{A Physical Example}
Here we present the physical example \cite{henara} which
actually initiated this work.
Phenomenological mean fields which contain quadrupole and octupole
deformations have been investigated classically \cite{HeNa}
and quantum mechanically \cite{ba,henara} as
this is of interest in nuclear physics and more recently also for the
description of metallic clusters. As a detailed discussion is presented in
the quoted papers we here focus our attention on the aspect of interest in
the present context.

The single particle potential
\begin{equation}
V(\varrho ,z)={m\over 2} \omega ^2
(\varrho ^2+{z^2\over b^2}+\lambda {2z^3-3z
\varrho ^2 \over
\sqrt{\varrho ^2+z^2}})
\label{p}
\end{equation}
is, for $b>1$ ($b<1$), a quadrupole deformed harmonic oscillator of
prolate (oblate) shape with an additional octupole term; in fact the term
multiplying $\lambda $ is proportional to $r^2P_3(\cos \theta )$ with $P_3$ the
third order Legendre polynomial. We use cylindrical coordinates $z$ and
$\varrho
=\sqrt{x^2+y^2} $. For $\lambda \ne 0$ this is a two degrees of freedom system
which is non-integrable. For $b<1$ it turns out that the switching on of the
octupole term very quickly gives rise to classically chaotic behaviour while
the onset of chaos is barely discernable when $b\ge 2$. The statistical
analyses of the correpsonding quantum spectra reveal the expected results in
that for $b<1$ the Wigner surmise is obtained for the nearest neighbour
distribution while for $b\ge 2$ a Poisson distribution is obtained, and for
particular values of $\lambda $ even a new shell structure emerges. The latter
is understood in terms of corresponding classical periodic orbits
\cite{ba,henara}.

In the spirit of the present paper the quantum mechanical findings should
be directly obtained from the matrix structure of the associated Hamiltonian
and the exceptional points related to it. The appropriate basis where
$H^0=p^2/2m+m\omega ^2
(\varrho ^2+z^2/b^2)/2$ is diagonal is given by the occupation
numbers $n_{\perp }$ and $n_z$. Note that the $z$-component of the angular
momentum is conserved and we consider here only $l_z=0$. The arrangement of
the quantum numbers in the matrix $H^1_{(n_z,n_{\perp }),(n_z',n_{\perp }')}$
is determined by the arithmetic ascending order of the unperturbed levels
$E^0_{n_z,n_{\perp }}$. The arrangement will therefore depend on the value of
$b$. The selection rules restrict entries in $H^1$ to $n_z'=n_z\pm (2k+1)$
and $n_{\perp }'=n_{\perp }\pm (2k)$ with $k=0,1,2,\ldots$.
It turns out that for $b\ge 2$ the entries vanish in the blocks of
$H^1$ which refer to the subspaces in which the unperturbed energies $E^0$
are degenerate. As a consequence, the perturbative expansion at $\lambda =0$
of the eigenvalues starts with the quadratic term. From the discussion in the
previous section it follows that a great number of exceptional points
is trapped at $\lambda =0$ and we therefore expect the onset of quantum chaos
only for values of $\lambda $ at an appreciable distance from zero. Using
$544\times 544$ matrices we illustrate in Fig.(4a) the distribution of the
real parts of the
exceptional points for the matrix problem $H^0+\lambda H^1$ with $b=2$. The
bulk
of the exceptional points occur in fact at values of $\lambda $ which fall
outside
the physical range (for $\lambda /\lambda _c\ge 1$ the potential no longer
binds). In
Fig.(4b) we diplay the corresponding distribution for $b=1/2$. Now we obtain
the maximum density of exceptional points as soon as the parameter
$\lambda $ is
switched on. This is in accordance with the results of the previous section
since now the linear term does occur in the expansion of each level
around $\lambda =0$ as there are non-zero entries in the blocks of $H^1$
which refer to the degenerate subspaces. The effect of this difference in
the distribution of the exceptional points manifests itself in the different
behaviour of the respective quantum spectra in that a Wigner distribution
is found for $\lambda >0$ when $b=1/2$ while for $b=2$ a Wigner distribution
never develops for $\lambda <\lambda _c$.

\section{Summary and Discussion}
The previous section provides a fine example where the statistical properties
of the quantum spectrum can be predicted from the properties of the individual
matrices $H^0$ and $H^1$ alone. To obtain the distribution of the exceptional
points as illustrated in Figs.(4) it is not necessary to solve the full
problem $H^0+\lambda H^1$ let alone to determine their positions (which would
be
impossible for more than 295,000 exceptional points). The distribution was
found using the method developed in \cite{hesan,kohe} and explained in
the Appendix. There, only some simple properties of $H^0$ and $H^1$ are used.
While this result alone yields just another confirmation of the method
employed in \cite{kohe}, the new aspect of this paper lies in the prediction
that, if the unperturbed levels are degenerate to second order,
level statistics ascribed to
quantum chaos are substantially suppressed initially and become manifest only
for sufficiently large values of the perturbation. The argument comes about
rather naturally from the behaviour of the exceptional points which in turn
determine the degree of quantum chaos \cite{hesan}.
While the previous section provides the physical relevance of our findings,
their universal character is argued and demonstrated in section two
and three. To the best of
our knowledge, this is the first example where even finer details with
regard to level statistics can be extracted from the distribution and the
general behaviour of the exceptional points. We believe that more refinement
can eventually even predict new shell structure as the one discovered in
the model discussed. Work towards this aim is in progress.

\appendix
\section{Distribution of the real parts of the exceptional points}
We use the method of the unperturbed curves \cite{kohe}, where the actual
spectrum is approximated by simple algebraic curves. The intesection points
of the approximate curves are then identified with the real parts of the
exceptional points. Since the aim is to obtain the distribution rather than
the excat positions, the approximate curves suffice. They are obtained from
the individual matrices $H^0$ and $H^1$ only. The requirement that the actual
spectrum and the approximate curves coincide exactly for small and for large
values of $\lambda $ leads for the prolate case ($b=2$) to the form
$$F_i(\lambda )=\left\{ \matrix{\varepsilon _i+\gamma _i \lambda ^2 \hfill &
            {\rm for}\  \lambda \le 0.9\lambda _c \cr  & \cr
  \root 3 \of{\omega ^3_i
  \lambda ^3+3\omega ^2_i \alpha _i\lambda ^2+c_i\lambda +
  d_i} &
            {\rm for}\  \lambda \ge 0.9\lambda _c } \right. $$
where the $\varepsilon _i$ and $\omega _i$ are the eigenvalues of
$H^0$ and $H^1$,
respectively, and
$$ \gamma _i= \sum_{n\ne i} {|H^1_{i,n}|^2\over \varepsilon _i-
\varepsilon _n},
\qquad
\alpha _i=(UH^0U^{-1})_{i,i} $$
with $U$ being the orthogonal matrix that diagonalises $H^1$. The coefficients
$c_i$ and $d_i$ are determined so as to smoothly match the two curves at
$\lambda =0.9\lambda _c$. We note that the second order perturbation
is correct up to
third order terms in $\lambda $ and is in fact very near to the actual spectrum
for $\lambda \le0.9\lambda _c$, while the expression for $\lambda \ge 0.9
\lambda _c$ is correct
up to terms of the order $1/\lambda $.

For the oblate case ($b=1/2$) the spectrum is, for the purpose considered,
sufficiently well approximated by the unperturbed lines
$$G_i(\lambda )=\varepsilon _i+\lambda \omega _i.$$
Recall that the eigenvalues $\varepsilon _i$ and $\omega _i$
are different from those
in the prolate case.

\newpage
\centerline{{\bf Figure captions}}

\vspace{0.5 cm}
{\bf Fig.1}
Exceptional points of the 9-dimensional model in the complex $\lambda $-plane
for
$\mu =0$ (a) and $\mu =1/2$ (diamonds in (b)) and $\mu =1$ (crosses in (b)).
Note the difference of the scale in (a) and (b) and the motion away from the
origin when $\mu $ increases.

\vspace{0.5 cm}
{\bf Fig.2}
Spectra of the 9-dimensional model for $\mu =0$ (a) and $\mu =1$ (b). For
better illustration a partial view is given for $\mu =1/2$ (c) and $\mu=1$
(d). Note how the position of the avoided crossing not only moves outwards
but also becomes more pronounced with increasing value of $\mu $.

\vspace{0.5 cm}
{\bf Fig.3}
Sample averages of nearest neighbour distributions of 304 levels for $\mu =0$
(a) with an eightfold degeneracy at $\lambda =0$ and for $\mu =1$ (b); both
distributions are for $\lambda =0.2$.

\vspace{0.5 cm}
{\bf Fig.4}
Distribution of the real parts of the exceptional points for the prolate case
(a) and the oblate case (b).
\end{document}